\documentclass[journal,10pt]{IEEEtran}
\ifCLASSINFOpdf

\else

\fi

\usepackage[utf8]{inputenc}
\usepackage{url}
\usepackage{cite}
\usepackage{hyperref}
\usepackage{graphicx}
\usepackage[nodisplayskipstretch]{setspace}
\usepackage{amsfonts} 
\usepackage{lipsum}
\usepackage{amsmath}
\usepackage{dsfont}
\usepackage{amssymb}
\usepackage{bbm}
\usepackage{amsmath}
\DeclareMathOperator*{\argmax}{arg\,max}

\DeclareMathOperator*{\limisup}{lim\,sup}
\DeclareMathOperator*{\amin}{\mbox{min.}}
\usepackage[dvipsnames]{xcolor}
\usepackage{csquotes}
\usepackage[utf8]{inputenc}
\usepackage[english]{babel}
\usepackage{dsfont}
\usepackage{amsthm,amssymb}
\usepackage{bbm}

\usepackage{blkarray}

\usepackage{algorithmicx}
\usepackage{algpseudocode}
\usepackage[norelsize, linesnumbered, ruled, lined, boxed, commentsnumbered]{algorithm2e}
\usepackage{tikz}
\usetikzlibrary{shapes, arrows, positioning,fit,backgrounds,calc}
\usepackage{pgfplots}
\pgfplotsset{compat=1.16}

\usepackage[percent]{overpic}
\usepackage[linesnumbered,ruled]{algorithm2e}
\usepackage[nodisplayskipstretch]{setspace}

\title{}
\author{}
\date{}
\allowdisplaybreaks
\title{
Real-time Tracking System with Partially Coupled Sources
}
\author{Saeid Sadeghi Vilni, and Risto~Wichman% <-this % stops a space
\thanks{%This research has been financially supported by.\\
Saeid Sadeghi Vilni and Risto Wichman are with Department of Information and Communications Engineering (DICE), Aalto University, 02150 Espoo, Finland (e-mail: Saeid.SadeghiVilni@aalto.fi; risto.wichman@aalto.fi). }}
\PassOptionsToPackage{silent}{bar}
\begin{document}

\maketitle
%\vspace{-10mm}
\begin{abstract}
    We consider a pull-based real-time tracking system consisting of multiple partially coupled sources and a sink. The sink monitors the sources in real-time and can request one source for an update at each time instant. The sources send updates over an unreliable wireless channel. The sources are partially coupled, and updates about one source can provide partial knowledge about other sources. We study the problem of minimizing the sum of an average distortion function and a transmission cost. Since the controller is at the sink side, the controller (sink) has only partial knowledge about the source states, and thus, we model the problem as a partially observable Markov decision process (POMDP) and then cast it as a belief-MDP problem. Using the relative value iteration algorithm, we solve the problem and propose a control policy. Simulation results show the proposed policy's effectiveness and superiority compared to a baseline policy.
    %We consider a pull-based real-time tracking system consisting of multiple partially coupled sources and a sink. Upon a request, a source sends an update over an unreliable channel. We study the problem of minimizing the sum of an average distortion function and a transmission cost. Since the sink has only partial knowledge about the source states, we model the problem as a partially observable Markov decision process (POMDP), cast it as a belief-MDP problem, and solve it via the relative value iteration algorithm to propose a control policy. Simulation results show the proposed policy's effectiveness and superiority compared to a baseline policy.
    
    \textbf{Index Terms:} Real-time tracking, pull-based system, coupled sources, partially observable Markov decision process. 
\end{abstract}

\section{Introduction}
Timely status updates are critical for real-time applications such as industrial automation, intelligent transportation, and smart grids. In these systems, a sensor observes a physical process such as industrial machine operations, vehicle movements, or power grid parameters and transmits a status update to an intended destination, such as a remote controller or a decision-making entity \cite{uysal2022semantic}. The effectiveness of real-time tracking depends not only on the update rate but also on the value and relevance of the received information. To capture this goal-oriented aspect, i.e., real-time tracking, distortion measure is a commonly used metric that captures the discrepancy between the sources' states and their estimate at the destination \cite{ssvrealtime}.

Recent studies on real-time tracking systems often assume independent sources. However, in many practical systems, the physical processes being monitored are partially coupled due to shared environments, network constraints, or system dependencies \cite{tong2022age, kalor2019minimizing}. This coupling introduces opportunities for more efficient update policies, as selecting one source for transmission can provide partial information about other sources, potentially reducing the overall estimation error and redundant transmissions. This motivates us to study a real-time tracking system with \textit{partially coupled} sources.

In this work, we consider a pull-based status update system where the sink selectively requests a single source to send an update. Upon receiving the command, the designated source transmits an update to the sink over an unreliable wireless channel (see Fig. \ref{fig:system model}). The underlying physical processes of the sources are partially coupled. More precisely, the partially coupled model is derived by combining fully coupled and independent models via a coupling factor. The considered system model may represent a status update system in a smart grid, where sensors monitor critical parameters like voltage, current, and temperature at different grid locations. These sensors are often coupled due to shared environmental factors and network inter-dependencies \cite{ma2023apriori}. {For instance, in power grids, voltage fluctuations at one bus can directly impact neighboring buses, leading to intrinsic coupling among voltage sensors.} The data is transmitted to a central control system, which tracks the grid’s status in real-time \cite{ma2023apriori}.

We address a real-time tracking problem, aiming to minimize the expected time-average sum of a distortion function and a transmission cost\footnote{The transmission cost represents the usage of radio resources.}. The solution to the problem defines a control policy that determines whether to stay idle or request an update and, if updating, selects the source to transmit. Since the state of the sources is only partially observable at the sink side, we model our problem as a partially observable Markov decision process (POMDP), and subsequently, we cast it as a belief-MDP problem. After truncating the belief-state space of the belief-MDP, we implement the relative value iteration algorithm (RVIA) to derive a control policy for solving the problem. The numerical results show the effectiveness of the proposed policy compared to a baseline policy.

Works \cite{tong2022age, kalor2019minimizing, hoang2021age, tripathi2022optimizing} investigate status update systems with correlated sources, focusing on age of information (AoI), while works \cite{erbayat2024age,ramakanth2024monitoring,abolfazlcorrelated} investigate estimation error in such systems.
The work \cite{erbayat2024age} derived closed-form expressions of average AoI and state estimation error in a multi-sensor system with correlated sources under an M/M/1/1 queue model. 
The work \cite{ramakanth2024monitoring} derived upper and lower bounds on estimation error using weighted sum AoI and proposed near-optimal AoI-based scheduling for correlated sources.
Unlike these works, we consider a pull-based status update system where the scheduler (sink) only has partial knowledge about the source states. 
In contrast to \cite{erbayat2024age,ramakanth2024monitoring}, we characterize the correlation through a coupling factor, which allows for a structured analysis of its impact.
The work in \cite{abolfazlcorrelated} studied a pull-based status update system with two correlated sources. Unlike our model, where the destination has only partial knowledge about the other sources, the correlation in \cite{abolfazlcorrelated} is captured through probabilistic observation. Specifically, the destination may decode the status of one source upon receiving an update from the other. Due to this fundamental difference, the policies derived in \cite{abolfazlcorrelated} are not applicable to our system. 
%In contrast to \cite{erbayat2024age,ramakanth2024monitoring}, which model correlation through predefined structures; 1) \cite{ramakanth2024monitoring} employed a discrete-time Wiener process with a general covariance matrix to model correlation and 2) \cite{erbayat2024age} used a general correlation matrix, our work introduces a partially coupled model. This approach integrates fully coupled and independent models using a coupling factor, offering a more flexible correlation model. 
Moreover, we solve an optimization control problem to minimize estimation error and transmission cost jointly, whereas \cite{erbayat2024age} characterized
a closed-form expression of the average estimation error and \cite{ramakanth2024monitoring} does not consider transmission cost.

\section{System Model and Problem Formulation}\label{sysm}
\subsection{System Model}\label{sysmA}
\begin{figure}
    \centering
    \includegraphics[width=8cm]{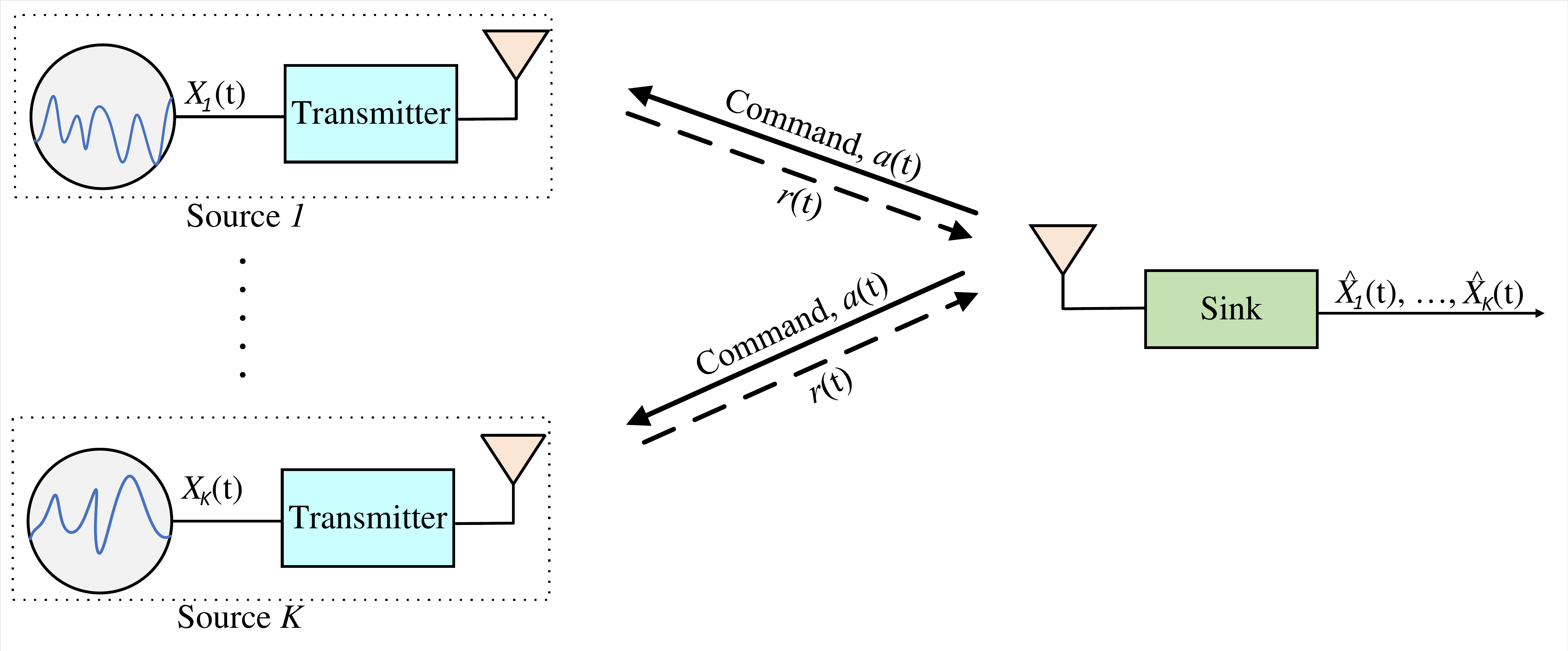}
    \caption{The considered system model.}
    \label{fig:system model}
\end{figure}

We consider a real-time tracking system consisting of $K$ sources with corresponding transmitters and a sink, as shown in Fig. \ref{fig:system model}. The sources are partially coupled, and each source is modeled as a binary time-varying discrete process. The sink is interested in the real-time tracking of the sources and sends a request  to a single source for an update at each time instant. Each source, upon receiving a command, sends an update. The time is discrete with unit time slots ${t \in \{1, 2, . . .\}}$.

\subsubsection{Source Model}%
To characterize the coupling between sources, we define independent sources and (fully) coupled sources, and then define partially coupled sources \cite{kennedy1992identification}.
\paragraph{Independent Source}% 
Let ${X_k(t)\in\{0,1\}}$ denote the state of source $k$ at slot $t$. Each source, at each slot, either remains in the same state with a probability $p$ or transits to any other state with a probability $ q = 1-p$.

Let $\textbf{X}(t)\in\mathcal{X}$ denote a vector containing all source states at slot $t$, i.e., $\textbf{X}(t)=[X_1(t),\dots,X_K(t)]$, where $\mathcal{X}$ is the space of all possible vectors of $\textbf{X}$\footnote{The space $\mathcal{X}$ is given as $\mathcal{X}=\{[j_1,\dots,j_K]\mid j_1,\dots,j_K\in\{0,1\}\}$}. Let $P^I(\textbf{X},\textbf{X}')$ denote the state transition probability of independent sources defined as the probability of moving from source states $\textbf{X}$ to source states $\textbf{X}'$. Since the sources are independent $P^I\big(\textbf{X},\textbf{X}')$ is given as
\begin{align}   P^I\big(\textbf{X},\textbf{X}')=\Pi_{k=1}^{k=K}\mathrm{Pr}(X_k'\mid X_k).
\end{align}
\paragraph{Coupled Sources} In the (fully) coupled model, the values of the sources are the same and move to the same state. If they start from different states, they become the same in the next slot. Let $P^C(\textbf{X},\textbf{X}')$ denote the state transition probability of coupled sources. Let \( \mathbf{1}_K \) and \( \mathbf{0}_K \) denote \( K \)-dimensional vectors with all elements equal to $1$ and $0$, respectively.
Thus, the state transition probability of coupled sources $P^C(\textbf{X},\textbf{X}')$ is given as
\begin{align}
   P^C(\textbf{X},\textbf{X}')\begin{cases}
       p& \textbf{X}=\mathbf{0}_K, \textbf{X}'=\mathbf{0}_K\\
       q& \textbf{X}=\mathbf{0}_K, \textbf{X}'=\mathbf{1}_K\\
       p& \textbf{X}=\mathbf{1}_K, \textbf{X}'=\mathbf{1}_K\\
       q& \textbf{X}=\mathbf{1}_K, \textbf{X}'=\mathbf{0}_K\\
       \theta& \textbf{X}\notin\{\mathbf{0}_K,\mathbf{1}_K\}, \textbf{X}'=\mathbf{0}_K\\
       1-\theta& \textbf{X}\notin\{\mathbf{0}_K,\mathbf{1}_K\}, \textbf{X}'=\mathbf{1}_K\\
       0&\text{otherwise},
   \end{cases} 
\end{align}
where $\theta$ determines the behavior of the sources when they start from different states.
\paragraph{Partially Coupled Sources} In the partially coupled model, the sources are not independent nor fully coupled. Let $P$ denote the state transition probability of partially coupled sources, given as
\begin{align}\label{P}
    P=\lambda P^C + (1-\lambda)P^I,
\end{align}
where $\lambda$ is the coupling factor.

\subsubsection{Status update} Only one channel is allocated to the system, and thus, one status update can happen at each slot. At the beginning of each slot, the sink decides whether to send a command to the transmitter or stay idle. Let $a(t)\in\{0,1,\dots,K\}$ denote the sink decision at slot $t$, where ${a(t)=k}$ indicates that the sink sends a command to source $k$, and $a(t)=0$ indicate that the sink stays idle, i.e., no command. We assume the control channel is error-free and instantaneous. %We denote the value of the received update at slot $t$ by $r(t)\in\{\sim,0,1\}$, where $r(t)=\sim$ indicates that no update was received.

Each source, upon receiving a command, sends a status update to the sink over an error-prone wireless channel. Let $r(t)\in\{0,1,\sim\}$ denote the value of the received update at slot $t$, where $r(t)=0$ means the value of the received update is $0$, $r(t)=1$ means the value of the received update is $1$, and $r(t)=\sim$ indicates that the sent updated is not delivered or the transmitter did not send an update. We denote the probability of successful delivery by $p_s$, i.e., $p_s=\mathrm{Pr}(a\neq 0,r\neq \sim)$.

\subsubsection{Estimation Strategy} \label{ES}
We assume the sink uses the maximum likelihood method for estimation. When the sink receives an update from a source, it knows the exact value of the source at the previous slot. However, for the current slot and subsequent slots without updates, we need to compute the probability distribution of the sources' states. For each source, the state with the highest probability at a given slot is interpreted as the value of that source. Let $\hat{X}_k(t)$ denote the estimate about the state of source $k$ at slot $t$. Let $\Phi(t) = [\phi_1(t),\dots,\phi_K(t)]$ denote the probability distribution of the source's states, where $\phi_k(t)$ is the probability distribution of source $k$ given as
\[
\phi_k(t)=\big[\mathrm{Pr}(X_k(t) = 0),\mathrm{Pr}(X_k(t) = 1)\big].
\]
Thus, the estimation at each slot is given as
\[
\hat{X}_k(t)=\argmax\{\phi_k(t)\}.
\]

The probability distribution of the source's states is a function of taken action at the previous slot, the value of the last received update at the previous slot, the state transition probability of the sources $P$, and the probability distribution at the previous slot $\Phi(t-1)$, i.e., ${\Phi(t)=f\big(a(t-1),r(t-1), P, \Phi(t-1)\big)}$. We assume that the initial distribution $\Phi(0)$ is known. 
%Given $\phi_k(0)$ and the state transition probability of the sources $P$, we can compute the probability distribution at each slot. 
The details regarding the calculation of $\Phi(t)$ are presented in the next section.

\subsection{Problem Formulation}\label{sysmB}
Our main goal is to jointly minimize the average estimation error and transmission cost by finding the optimal action $a(t)$ at each slot. At each slot, the action is taken based on available information about the sources' states and their estimation at the sink.

The estimation error, defined as the discrepancy between the source state $X_k$ and its estimate at the sink $\hat{X}_k$, is quantified via a distortion function. 
Let $D: \big(X,\hat X\big ) \rightarrow \mathbb{R}^+$ denote a (generic) bounded distortion function, i.e., $|D(.)|<\infty$. For instance, the distortion function can be defined as the error indicator, i.e., ${D\big(X,\hat X\big )=\mathbbm{1}_{\{X\neq X\}} }$\footnote{$\mathds{1}_{\{.\}}$ is an indicator function that equals $1$ when the condition(s) in $\{.\}$ is satisfied.}, the absolute value of the difference between $X(t)$ and $\hat{X}(t)$, i.e., ${D\big(X,\hat X\big )= | X-\hat X |}$, or mean squared error, i.e., ${D\big(X,\hat X\big )= | X-\hat X |^2}$. 

Let $\tilde{c}(t)$ denote the cost for a transmission attempt for the sources at slot $t$, given as
\begin{align}
    \tilde{c}(t)=\mathbbm{1}_{\{a(t)\neq 0\}}.
\end{align}

Formally, the considered problem can be formulated as the following stochastic control problem:
\begin{subequations}\label{pmain}
\begin{alignat}{2}
\displaystyle\amin_{a(t)} ~& \underset{T\rightarrow \infty}{\limisup} ~ \frac{1}{T} \sum_{t=1}^{T}\mathbb{E} \{ \frac{1}{K} \sum_{k=1}^{K}D\big(X_k(t),\hat X_k(t)\big ) + \gamma\tilde{c}(t)\}\\
\mbox{s.t.}~ 
&a(t)\in \{0,\dots,K\}\label{p1:1},
\end{alignat}
\end{subequations}
where $\{a(t)\}_{t\in\{1,2, \dots\}}$ are the optimization variables, and $\gamma$ is the weight of the transmission cost. The expression \eqref{p1:1} indicates the binary nature of the decision variable. In problem \eqref{pmain}, $\mathbb{E}\{\cdot\}$ is the expectation with respect to the randomness of the system (i.e., randomness in the source dynamic and the forward communication channels) and the decision variable $a(t)$.
\section{Transmission Scheduling Policy}\label{policy}
In this section, we propose a control policy to solve problem \eqref{pmain}. Since the controller is at the sink side and the forward channel is error-pron, the state of the source is only partially observable at the sink side. Therefore, we model problem \eqref{pmain} as a partially observable Markov decision process (POMDP). Then, we cast the POMDP problem into a belief-MDP problem and solve it via the relative value iteration algorithm (RVIA).

\subsection{POMDP Formulation}
A POMDP has five elements, which are elaborated below:\\
\textbf{State}: Let $s(t)=\{\textbf{X}(t),r(t-1),\Phi(t)\}\in\mathcal{S}$ denote the state of the system at slot $t$, where 
%$\textbf{X}(t)=\{X_1(t),\dots,X_K(t)\}$, $\hat{\textbf{X}}=\{\hat X_1(t),\dots,\hat X_K(t)\}$, and 
$\mathcal{S}$ is the state space.\\
\textbf{Action}: The action is the same as the command $a(t)$.\\
\textbf{Cost}: Let $C(t)$ denote the immediate cost at slot $t$ given as ${C(t)= \frac{1}{K}\sum_{k=1}^{K}D\big(X_k(t),\hat X_k(t)\big ) + \gamma\tilde{c}(t)}$.\\
\textbf{Observation}: Let $o(t)$ denote the observation at slot $t$ defined as the observable part of the state, given as ${o(t)=\{r(t-1),\Phi(t)\}}$. \\
\textbf{State transition probability}: Let $\mathcal{P}(s,s',a)$ denote the state transition probability defined as the probability of moving from state $s$ to state $s'$ given action $a$. We denote $\bar{p}_s=1-p_s$ and use $r_{-1}$ to denote the value of the received update at the previous slot, $s=\{{\textbf{X}}, r_{-1},\Phi\}$, and $s'=\{{\textbf{X}'},r_{-1}',\Phi'\}$. Since the sources' dynamic is independent of the other elements of the state, the state transition probability is given as
\begin{align}
    \mathcal{P}(s,s',a)=\mathrm{Pr}(\textbf{X}'\mid \textbf{X})\mathrm{Pr}(r_{-1}',\Phi'\mid s),
\end{align}
where $\mathrm{Pr}(\textbf{X}'\mid \textbf{X})$ is derived directly from $P$ (presented in \eqref{P}). The expression $\mathrm{Pr}(r_{-1}',\Phi'\mid s)$ is derived as
\begin{align*}
    &\mathrm{Pr}(r_{-1}',\Phi'\mid s)=\notag\\&\begin{cases} \phi_{k,0}p_s&a=k,~r'_{-1}=0,~\Phi'=f\big(k,1,0,P, \Phi\big),\\ \phi_{k,1}p_s&a=k,~r'_{-1}=1,~\Phi'=f\big(k,1,1,P, \Phi\big),\\ \bar{p}_s&a=k,~r'_{-1}=\sim,~\Phi'=f\big(k,0,\sim,P, \Phi\big),\\ 1&a=0,~r'_{-1}=\sim,~\Phi'=f\big(0,0,\sim,P, \Phi\big).
    \end{cases}
\end{align*}
\subsection{Belief-MDP Characterization}
The state of the sources $\textbf{X}$ is only partially observable at the controller (sink) for decision-making. To address this insufficiency, we utilize the notion of \textit{belief} \cite[Chapter 7]{sigaud2013markov}, and we cast the above-defined POMDP as a belief-MDP. To this end, we define a belief about the source states that retain the Markov property and capture all information necessary for decision-making.

Let $I^c(t)$ denote the complete information state at slot $t$, which consists of: 1) initial probability distribution over the states, 2) the history of past and current observations, i.e., ${\{o(1),\dots,o(t)\}}$, and 3) the history of past actions, i.e., ${\{a(1),\dots,a(t-1)\}}$. Let ${\mathcal{B}(t)=\{\boldsymbol{b}_{1}(t),\dots,\boldsymbol{b}_{K}(t)\}}$ denote a set of beliefs about the source states at slot $t$, where $\boldsymbol{b}_{k}(t)$ is a vector ${\boldsymbol{b}_{k}(t)= [b_{k,0}(t),b_{k,1}(t)]}$ and the entries denote the conditional probabilities of the possible values of $X(t)$ given $I^c(t)$, i.e., $b_{k,i}(t)\triangleq\mathrm{Pr}\big(X_k(t)=i\mid I^c(t)\big)$ and $i\in\{0,1\}$. Note that $\phi_k(t)$ presented in \ref{ES} is derived directly by $\boldsymbol{b}_{k}(t)$.

The belief at slot $t+1$, $\boldsymbol{b}_{k}(t+1)$, is updated based on the previous belief $\boldsymbol{b}_{k}(t)$, the current observation $o(t+1)$, and the previous action $a(t)$. The belief updates are derived for two possible cases, namely: 1) Case $1$, i.e., $r(t)=\sim$, and 2) Case $2$, i.e., $a(t)=u$, $r(t)=m$ where $m\in\{0,1\}$. The belief update is expressed as
\begin{align}\label{beliefupdate1}
    &{b}_{k,i}(t+1) =\notag\\ 
    &\begin{cases}
        \displaystyle\sum_{\hat{\textbf{X}} \in \mathcal{X}} \sum_{\tilde{\textbf{X}} \in {\tilde{\mathcal{X}}}} 
        \mathrm{Pr}\big(\textbf{X}(t)=\hat{\textbf{X}}\mid I^c(t)\big) P(\hat{\textbf{X}}, \tilde{\textbf{X}}), &\text{Case}~1,\\
        \displaystyle\sum_{\hat{\textbf{X}} \in {\mathcal{X}}^u} \sum_{\tilde{\textbf{X}} \in {\tilde{\mathcal{X}}}} 
        \mathrm{Pr}\big(\textbf{X}(t)=\hat{\textbf{X}}\mid I^c(t)\big) P(\hat{\textbf{X}},\tilde{\textbf{X}}), & \text{Case}~2,
    \end{cases}
\end{align}
where ${\tilde{\mathcal{X}}}=\{[j_1,\dots,j_K]\mid j_1,\dots,j_K\in\{0,1\}, j_k=i\}$, and ${{\mathcal{X}}^u}=\{[j_1,\dots,j_K]\mid j_1,\dots,j_K\in\{0,1\},j_u=m\}$.

Let $\textbf{b}_{\hat{\textbf{X}}}(t)$ denote a belief about the joint values of sources' state at slot $t$, i.e., $\textbf{b}_{\hat{\textbf{X}}}(t)=\mathrm{Pr}\big(\textbf{X}(t)=\hat{\textbf{X}}\mid I^c(t)\big)$.
The evolution of $\textbf{b}_{\hat{\textbf{X}}}(t)$ is given as
\begin{align}\label{jointbeliefupdate} 
&\textbf{b}_{\hat{\textbf{X}}}(t+1)=\begin{cases}
        \displaystyle\sum_{\bar{\textbf{X}} \in \mathcal{X}}\textbf{b}_{\bar{\textbf{X}}}(t)P(\bar{\textbf{X}},\hat{\textbf{X}}), & \text{Case}~1,\\
        \displaystyle\sum_{\bar{\textbf{X}} \in {\mathcal{X}}^u} 
        \frac{\textbf{b}_{\bar{\textbf{X}}}(t)}{\sum_{\bar{\textbf{X}} \in {\mathcal{X}}^u}\textbf{b}_{\bar{\textbf{X}}}(t)} P(\bar{\textbf{X}},\hat{\textbf{X}}), & \text{Case}~2,
    \end{cases}
\end{align}
where Case 2 follows the fact that, we have the partial information that $\bar{\textbf{X}} \in {\mathcal{X}}^u$ and thus $\textbf{b}_{\bar{\textbf{X}}}=0~\forall~\bar{\textbf{X}} \notin {\mathcal{X}}^u$. Thus, we need to normalize the belief values such that they sum to one within the restricted set ${\mathcal{X}}^u$.

%Given $\textbf{b}_{\hat{\textbf{X}}}(t)=\mathrm{Pr}\big(\textbf{X}(t)=\hat{\textbf{X}}\mid I^c(t)\big)$, 
By substituting \eqref{jointbeliefupdate} in \eqref{beliefupdate1}, the belief update is derived as
\begin{align}\label{beliefupdate2}
    &{b}_{k,i}(t+1) =\begin{cases}
        \displaystyle\sum_{\hat{\textbf{X}} \in \mathcal{X}} \sum_{\tilde{\textbf{X}} \in {\tilde{\mathcal{X}}}} 
        \textbf{b}_{\hat{\textbf{X}}}(t) P(\hat{\textbf{X}},\tilde{\textbf{X}}), & \text{Case}~1,\\
        \displaystyle\sum_{\hat{\textbf{X}} \in {{\mathcal{X}}^u}} \sum_{\tilde{\textbf{X}} \in {\tilde{\mathcal{X}}}} 
        \frac{\textbf{b}_{{\hat{\textbf{X}}}}(t)}{\sum_{{\hat{\textbf{X}}} \in {\mathcal{X}}^u}\textbf{b}_{{\hat{\textbf{X}}}}(t)}  P(\hat{\textbf{X}},\tilde{\textbf{X}}), & \text{Case}~2.
    \end{cases}
\end{align}

Given the belief, the belief-state at slot $t$ is defined as ${z(t)=\{\mathcal{B}(t),r(t-1)\}\in\mathcal{Z}}$, where $\mathcal{Z}$ is the belief-state space. Note that since $\Phi(t)$ is derived directly from $\mathcal{B}(t)$, it is excluded from the belief-state.

To derive the belief-state transition probabilities, we denote the next belief via $\hat{\mathcal{B}}$ if it is derived by the first case in \eqref{beliefupdate2} and via $\tilde{\mathcal{B}}$ if it is derived by the second case in \eqref{beliefupdate2}. Thus, the belief-state transition probabilities for ${z=\{\mathcal{B},r_{-1}\}}$ and different actions are given as
\begin{align}
    &\mathrm{Pr}(z'\mid z,a)=\notag\\&\begin{cases}1,&a=0,~\mathcal{B}'=\hat{\mathcal{B}},~r_{-1}=\sim,\\\bar{p}_s,&a\neq0,~\mathcal{B}'=\hat{\mathcal{B}},~r_{-1}=\sim,\\b_{k,0}p_s,&a=k,~\mathcal{B}'=\tilde{\mathcal{B}},~r_{-1}=0,\\b_{k,1}p_s,&a=k,~\mathcal{B}'=\tilde{\mathcal{B}},~r_{-1}=1.
    \end{cases}
\end{align}
%Note that $\hat{{X}}_k'=\argmax\{\boldsymbol{b}_k'\}$ where $\boldsymbol{b}_k'\in \mathcal{B}'$.

Finally, The immediate cost for the belief-MDP is defined as the expected cost of POMDP $C(t)$, given as
\begin{align}
    \hat{C}(t)= \frac{1}{K}\sum_{k=1}^{K}\sum_{i=0}^1 b_{k,i}D\big(i,\hat X_k(t)\big ) + \gamma\tilde{c}(t).
\end{align}

Thus, the Belief-MDP problem is given as
\begin{subequations}\label{pbmdp}
\begin{alignat}{2}
\displaystyle\amin_{\pi} \quad & \underset{T\rightarrow \infty}{\limisup} ~ \frac{1}{T}\sum_{t=1}^{T}\mathbb{E} \{\hat{C}(t) \}.
\end{alignat}
\end{subequations}
where $\pi$ is the control policy.

To solve the belief-MDP problem, we implement the relative value iteration algorithm (RVIA)\footnote{For RVIA convergence, the MDP must be unichain and aperiodic \cite[Proposition 4.3.2]{bertks}, which is nontrivial to verify. However, numerical results indicate RVI convergence.} \cite[Proposition 4.3.2]{bertks}.
The RVIA requires a finite belief-state and action space. However, due to the belief element, the belief-state space is infinite. To truncate it, we define a finite space comprising belief trajectories over \(N\) transition steps for all action combinations. Beliefs at step \(N+1\) are projected onto the closest belief within this space, measured by mean-square error. The belief-state transition probabilities are then updated to account for transitions from step \(N\) to the projected beliefs at step \(N+1\). The derived policy is referred to as POMDP-based policy.

\section{Numerical Results}\label{snrs}
In this section, we evaluate the performance of the POMDP-based policy compared to a maximum age first (MAF) policy in which the source with the largest AoI is requested to send an update. In the simulation results, we use the following distortion function: ${ D\big(X,\hat{X}\big) = |X-\hat{X}|}$.

\begin{figure}[!ht]
\centering
\begin{tikzpicture}[scale=0.95]
\begin{axis}[
    title={},
    %caption = {aaaa},
    xlabel={Coupling factor, $\lambda$},
    ylabel={Average cost},
    xmin=0.1, xmax=0.9,
    ymin=0.45, ymax=0.57,
    xtick={0.1,0.3,0.5,0.7,0.9},
    xmajorgrids=true,
    grid style=dashed,
    ytick={0.45,0.48,0.51,0.54,0.57},
    legend pos=north east,
    ymajorgrids=true,
    grid style=dashed,
]

        \addplot[
    color=blue,
    mark=o,
    line width=1.3pt, % Increase the line width
    mark size=3pt,
    dashed,
    mark options={solid}
    ]
    coordinates {
    (0.9,0.4981)(0.7,0.5202)(0.5,0.5410)(0.3,0.5618)(0.1,0.5655)
    };

        \addplot[
    color=red,
    mark=diamond,
    line width=1.3pt, % Increase the line width
    mark size=3pt,
    dashed,
    mark options={solid}
    ]
    coordinates {
    (0.9,0.4928)(0.7,0.5070)(0.5,0.5222)(0.3,0.5369)(0.1,0.5389)
    };

        \addplot[
    color=blue,
    mark=o,
    line width=1.3pt, % Increase the line width
    mark size=3pt,
    ]
    coordinates {
    (0.9,0.4543)(0.7,0.4695)(0.5,0.4824)(0.3,0.4931)(0.1,0.5038)
    };

            \addplot[
    color=red,
    mark=diamond,
    line width=1.3pt, % Increase the line width
    mark size=3pt,
    ]
    coordinates {
    (0.9,0.4502)(0.7,0.4584)(0.5,0.4681)(0.3,0.4785)(0.1,0.4860)
    };

    \legend{MAF policy $K=3$,MAF policy $K=2$,POMDP-based policy $K=3$,POMDP-based policy $K=2$}
    
\end{axis}
\end{tikzpicture}
%\vspace{-5mm}
  \caption{The average cost of different policies with respect to coupling factor $\lambda$, where $p_s=0.8$, $p = 0.8$, $\gamma = 0.15$, $N = 6$.}\label{fig:lambda}
  \end{figure}
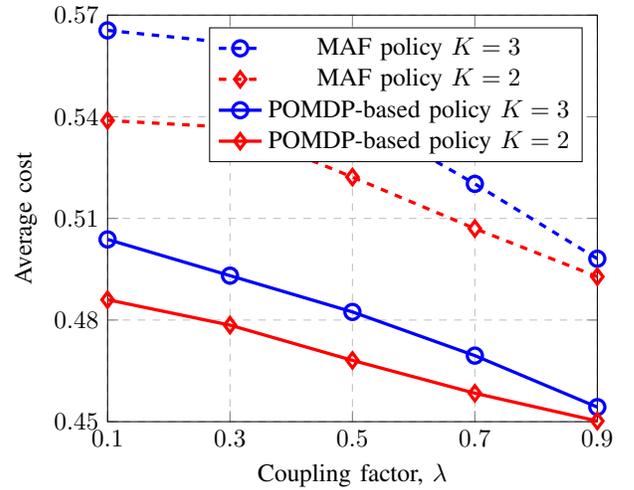

Fig. \ref{fig:lambda} shows the average cost of the POMDP-based and MAF policies for different numbers of sources versus the coupling factor $\lambda$. {As can be seen, the average cost decreases when the coupling factor increases. This is because, as the coupling between the sources increases, an update from one source provides more information about other sources' states.} Moreover, the POMDP-based policy outperforms the MAF policy. This is due to the fact that %the POMD-based policy commands effectively such that at some slots, the sink stays idle, 
with the POMD-based policy, the sink may stay idle at some slots, 
while by MAF policy,one source is requested to transmit  at each slot. 

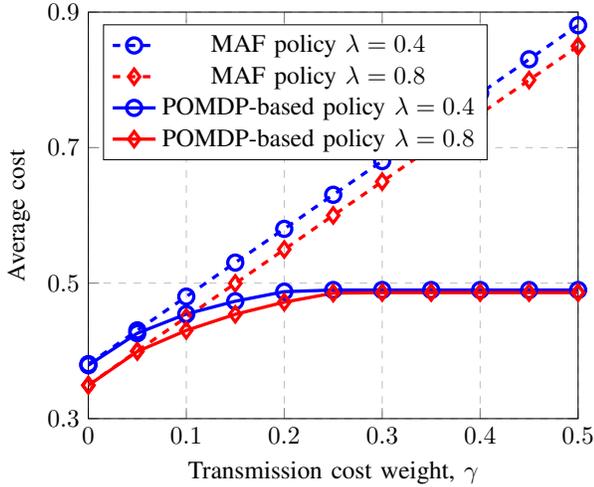
\begin{figure}[!ht]
\centering
\begin{tikzpicture}[scale=0.95]
\begin{axis}[
    title={},
    %caption = {aaaa},
    xlabel={Transmission cost weight, $\gamma$},
    ylabel={Average cost},
    xmin=0.0, xmax=0.5,
    ymin=0.3, ymax=0.9,
    xtick={0.0,0.1,0.2,0.3,0.4,0.5},
    xmajorgrids=true,
    grid style=dashed,
    ytick={0.3,0.5,0.7,0.9,1.1},
    legend pos=north west,
    ymajorgrids=true,
    grid style=dashed,
]

        \addplot[
    color=blue,
    mark=o,
    line width=1.3pt, % Increase the line width
    mark size=3pt,
    dashed,
    mark options={solid}
    ]
    coordinates {
    (0.0000,0.3805)(0.0500,0.4303)(0.1000,0.4801)(0.1500,0.5303)(0.2000,0.5801)(0.2500,0.6304)(0.3000,0.6801)(0.3500,0.7303)(0.4000,0.7801)(0.4500,0.8304)(0.5000,0.8806)

    };

        \addplot[
    color=red,
    mark=diamond,
    line width=1.3pt, % Increase the line width
    mark size=3pt,
    dashed,
    mark options={solid}
    ]
    coordinates {
    (0.0000,0.3495)(0.0500,0.4000)(0.1000,0.4496)(0.1500,0.4996)(0.2000,0.5497)(0.2500,0.6000)(0.3000,0.6498)(0.3500,0.6999)(0.4000,0.7498)(0.4500,0.7997)(0.5000,0.8497)(0.5000,0.8497)
    };

        \addplot[
    color=blue,
    mark=o,
    line width=1.3pt, % Increase the line width
    mark size=3pt,
    ]
    coordinates {
    (0.0000,0.3788)(0.0500,0.4260)(0.1000,0.4544)(0.1500,0.4733)(0.2000,0.4874)(0.2500,0.4898)(0.3000,0.4898)(0.3500,0.4898)(0.4000,0.4898)(0.4500,0.4898)(0.5000,0.4898)
    };

            \addplot[
    color=red,
    mark=diamond,
    line width=1.3pt, % Increase the line width
    mark size=3pt,
    ]
    coordinates {
    (0.0000,0.3491)(0.0500,0.3991)(0.1000,0.4302)(0.1500,0.4540)(0.2000,0.4719)(0.2500,0.4852)(0.3000,0.4860)(0.3500,0.4860)(0.4000,0.4860)(0.4500,0.4860)(0.5000,0.4860)(0.5000,0.4860)
    };

    \legend{MAF policy $\lambda=0.4$,MAF policy $\lambda=0.8$,POMDP-based policy $\lambda=0.4$,POMDP-based policy $\lambda=0.8$}
    
\end{axis}
\end{tikzpicture}
%\vspace{-5mm}
  \caption{The average cost of different policies with respect to transmission cost weight $\gamma$, where $p_s=0.8$, $p = 0.8$,  $K=2$, $N = 6$.}\label{fig:gamma}
  \end{figure}

Fig. \ref{fig:gamma} depicts the average cost of different policies for different coupling factors with respect to the transmission cost weight $\gamma$. The figure shows that the MAF policy increases with $\gamma$ linearly as expected, whereas the POMD-based policy, for $\gamma$ larger than $0.3$, does not change. This is because, for these values of $\gamma$, the policy tends to request updates infrequently; in other words, the transmission cost dominates the distortion cost promoting less transmission. 

\begin{figure}[!ht]
\centering
\begin{tikzpicture}[scale=0.95]
\begin{axis}[
    title={},
    %caption = {aaaa},
    xlabel={Probability of successful delivery, $p_s$},
    ylabel={Average cost},
    xmin=0.2, xmax=1,
    ymin=0.399, ymax=0.51,
    xtick={0.2,0.4,0.6,0.8,1},
    xmajorgrids=true,
    grid style=dashed,
    ytick={0.4,0.42,0.44,0.46,0.48,0.5},
    legend pos=north east,
    ymajorgrids=true,
    grid style=dashed,
]

        \addplot[
    color=blue,
    mark=o,
    line width=1.3pt, % Increase the line width
    mark size=3pt,
    dashed,
    mark options={solid}
    ]
    coordinates {
    (0.2,0.5003)(0.4,0.4684)(0.6,0.4462)(0.8,0.4305)(1,0.418)
    };

        \addplot[
    color=blue,
    mark=o,
    line width=1.3pt, % Increase the line width
    mark size=3pt,
    ]
    coordinates {
    (0.2,0.4936)(0.4,0.4632)(0.6,0.4431)(0.8,0.4288)(1,0.4169)
    };

        \addplot[
    color=red,
    mark=diamond,
    line width=1.3pt, % Increase the line width
    mark size=3pt,
    dashed,
    mark options={solid}
    ]
    coordinates {
    (0.2,0.4920)(0.4,0.4557)(0.6,0.4316)(0.8,0.4144)(1,0.4015)
    };

        \addplot[
    color=red,
    mark=diamond,
    line width=1.3pt, % Increase the line width
    mark size=3pt,
    ]
    coordinates {
    (0.2,0.4826)(0.4,0.4504)(0.6,0.4289)(0.8,0.4121)(1,0.3997)
    };

    \legend{MAF policy $K=3$,POMDP-based policy $K=3$,MAF policy $K=2$, POMDP-based policy$K=2$}
    
\end{axis}
\end{tikzpicture}
%\vspace{-5mm}
  \caption{The average cost of different policies with respect to the probability of successful delivery $p_s$, where $\lambda=0.6$, $p = 0.8$, $\gamma = 0.05$, $N = 6$.}\label{fig:channel}
  \end{figure}
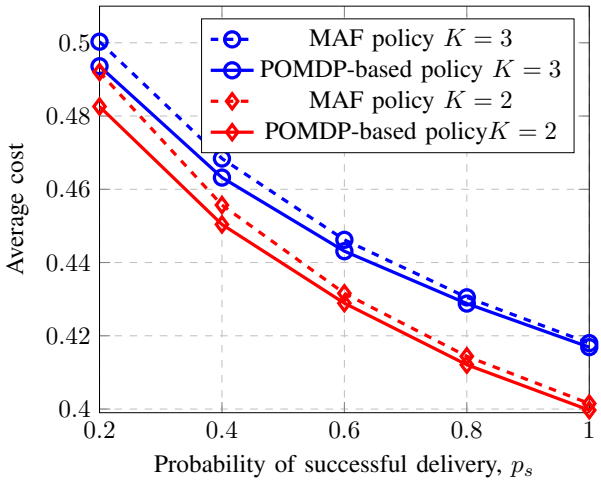

In Fig. \ref{fig:channel}, we examine the average cost of different policies for different numbers of sources with respect to the probability of successful delivery $p_s$. The figure shows that, as expected, the average cost decreases when $p_s$ increases. This is because the number of successful transmissions increases. Moreover, the increase in the number of sources, also shown in Fig. \ref{fig:lambda}, increases the average cost. This is because at each slot, only one source can be requested for the update, and thus, an increase in the number of sources increases 1) the number of requests which increases the transmission cost, and 2) the number of slots without an update for the sources which increases uncertainty.

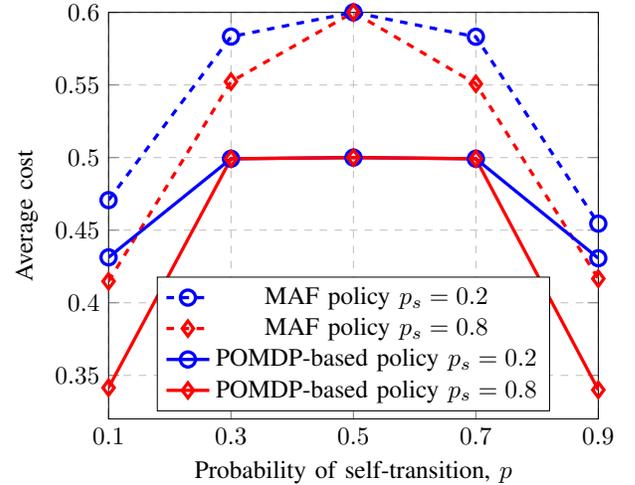
\begin{figure}[!ht]
\centering
\begin{tikzpicture}[scale=0.95]
\begin{axis}[
    title={},
    %caption = {aaaa},
    xlabel={Probability of self-transition, $p$},
    ylabel={Average cost},
    xmin=0.1, xmax=0.9,
    ymin=0.32, ymax=0.6,
    xtick={0.1,0.3,0.5,0.7,0.9},
    xmajorgrids=true,
    grid style=dashed,
    ytick={0.35,0.4,0.45,0.5,0.55,0.6},
    legend style={at={(0.5,0.35)}, anchor=north},
    ymajorgrids=true,
    grid style=dashed,
]

        \addplot[
    color=blue,
    mark=o,
    line width=1.3pt, % Increase the line width
    mark size=3pt,
    dashed,
    mark options={solid}
    ]
    coordinates {
    (0.1000,0.4707)(0.3000,0.5833)(0.5000,0.6000)(0.7000,0.5832)(0.9000,0.4545)

    };

        \addplot[
    color=red,
    mark=diamond,
    line width=1.3pt, % Increase the line width
    mark size=3pt,
    dashed,
    mark options={solid}
    ]
    coordinates {
    (0.1000,0.4148)(0.3000,0.5523)(0.5000,0.6000)(0.7000,0.5508)(0.9000,0.4166)
    };
        \addplot[
    color=blue,
    mark=o,
    line width=1.3pt, % Increase the line width
    mark size=3pt,
    ]
    coordinates {
    (0.1000,0.4313)(0.3000,0.4992)(0.5000,0.5000)(0.7000,0.4992)(0.9000,0.4307)
    };

            \addplot[
    color=red,
    mark=diamond,
    line width=1.3pt, % Increase the line width
    mark size=3pt,
    ]
    coordinates {
    (0.1000,0.3415)(0.3000,0.4992)(0.5000,0.5000)(0.7000,0.4992)(0.9000,0.34)
    };

    \legend{MAF policy $p_s=0.2$,MAF policy $p_s=0.8$,POMDP-based policy $p_s=0.2$,POMDP-based policy $p_s=0.8$}   
\end{axis}
\end{tikzpicture}
%\vspace{-5mm}
  \caption{The average cost of different policies for different probability of successful delivery with respect to the probability of self-transition $p$, where $\lambda = 0.4$, $K=2$, $\gamma=0.05$, $N = 6$.}\label{fig:p}
  \end{figure}

Fig. \ref{fig:p} demonstrates the average cost of different policies for different probabilities of successful delivery with respect to the probability of self-transition $p$. As can be seen, the average cost is small for small and large values of $p$, while for $p=0.5$, the average cost is the largest. This is because, for small and large values of $p$, the sources' next state can be estimated more accurately, which reduces the uncertainty and, subsequently, results in better estimates, while for $p=0.5$, the uncertainty about the source state is the highest which degrades the estimation quality. Moreover, the average cost of POMDP-based policy for $p=0.3$, $p=0.5$, and $p=0.7$ is quite the same for good and weak channel conditions. This is because for these values of $p$, the sources' dynamics is high, and there is a chance that the estimation error decreases by itself, which promotes less transmission.

\section{Conclusions}\label{clc}
We considered a pull-based real-time tracking system monitoring multiple partially coupled sources. Upon receiving a command from the sink, the source sends an update over an unreliable wireless channel. We addressed the problem of minimizing the average sum of a distortion function and a transmission cost. Since the sources are partially observable at the sink side, we modeled the problem as a POMDP problem, which is then cast as a belief-MDP problem. We truncated the belief-state space, implemented RVIA to solve the belief-MDP problem, and proposed a POMDP-based policy. The simulation results showed the effectiveness of the POMDP-based policy compared to the MAF policy. The results suggest that the average cost decreases with an increase in the coupling factor. Moreover, the results show that the policy promotes less transmission for high dynamic sources, i.e., $p=0.5$.
\appendices

\vspace{-4mm}
\bibliographystyle{IEEEtran}
%\begin{spacing}{1.35}
\bibliography{short-conf,short-jour,Main}
%\end{spacing}

\end{document}